 \documentclass[preprint2]{emulateapj}
\shortauthors{\"Oberg et al.}

\begin{document}

\title{Complex organic molecules during low-mass star formation:\\ Pilot survey results}

\author{Karin I. \"Oberg}
\affil{Harvard-Smithsonian Center for Astrophysics, 60 Garden Street, Cambridge, MA 02138, USA}
\email{koberg@cfa.harvard.edu}

\author{Trish Lauck}
\affil{University of Virginia, Charlottesville, VA 22904, USA}

\author{Dawn Graninger}
\affil{Harvard-Smithsonian Center for Astrophysics, 60 Garden Street, Cambridge, MA 02138, USA}

\begin{abstract}
Complex organic molecules (COMs) are known to be abundant toward {\it some} low-mass young stellar objects (YSOs), but how these detections relate to typical COM abundance are not yet understood. We aim to constrain the frequency distribution of COMs during low-mass star formation, beginning with this pilot survey of COM lines toward six embedded YSOs using the IRAM 30m telescope. The sample was selected from the Spitzer $c2d$ ice sample and covers a range of ice abundances. We detect multiple COMs, including CH$_3$CN, toward two of the YSOs, and tentatively toward a third. Abundances with respect to CH$_3$OH vary between 0.7 and 10\%. This sample is combined with previous COM observations  and upper limits to obtain a frequency distributions of CH$_3$CN, HCOOCH$_3$, CH$_3$OCH$_3$ and CH$_3$CHO. We find that for all molecules more than 50\% of the sample have detections or upper limits of 1--10\% with respect to CH$_3$OH. Moderate abundances of COMs thus appear common during the early stages of low-mass star formation. A larger sample is required, however, to quantify the COM distributions, as well as to constrain the origins of observed variations across the sample.\footnote{Based on observations carried out with the IRAM Plateau de Bure Interferometer. IRAM is supported by INSU/CNRS (France), MPG (Germany) and IGN (Spain).}
\end{abstract}


\section{Introduction}

\begin{table*}%[htp]
{\scriptsize
\begin{center}
\caption{Source information. \label{sources}}
\begin{tabular}{l c c c c c c c c c c}
\hline\hline
Source & R.A. & Dec & Cloud & L$_{\rm bol}^{\rm a,b,c}$	&M$_{\rm env}^{\rm a,b,c}$	&$\alpha_{\rm IR}^{\rm d}$&$N_{\rm H_2O}^{\rm d}$ & $x_{\rm CH_3OH}^{\rm d}$ & $x_{\rm NH_3}^{\rm e}$	&$x_{\rm CH_4}^{\rm f}$\\
	&(J2000.0)	&(J2000.0)		&&L$_\odot$	&M$_\odot$ & &10$^{18}$ cm$^{-2}$	&\%$N_{\rm H_2O}$	&\%$N_{\rm H_2O}$&\%$N_{\rm H_2O}$\\
\hline
IRAS 03235+3004	&03 26 37.45	&+30 15 27.9	&Perseus	&1.9	&2.4	&1.44&14[2]	&4.2[1.2]	&4.7[1.0]	&4.3[1.4]\\
B1-a	&03 33 16.67	&+31 07 55.1				&Perseus	&1.3	&2.8	&1.87&$<$1.9	&3.3[1.0]	&$<$5.8\\
{\it B1-b}	&03:33:20.34	&31:07:21.4			&Perseus	&--&26	&0.68&18[3]	&11.2[0.7]	&4.2[2.0]	&3.3[0.6]\\	
B5 IRS 1	&03 47 41.61	&+32 51 43.8			&Perseus	&4.7	&4.2	&0.78&2.3[0.3]	&$<$3.7	&$<$2.1	&--	\\
L1489 IRS	&04 04 43.07	&+26 18 56.4		&Taurus	&3.7	&--&1.10&4.3[0.5]	&4.9[1.5]	&5.4[1.0]	&3.1[0.2]\\
IRAS 04108+2803	&04 13 54.72	&+28 11 32.9	&Taurus	&0.62	&--&0.90&2.9[0.4]	&$<$3.5	&4.3[1.0]	&$<$10\\
SVS 4-5	&18 29 57.59	&+01 13 00.6			&Serpens	&38	&3.5	&1.26&5.7[1.1]	&25[4]	&4.3[2.0]	&6.1[1.7]	\\
\hline
\end{tabular}
\\$^{\rm a}$\citet{Pontoppidan04}, $^{\rm b}$\citet{Hatchell07}, $^{\rm c}$\citet{Furlan08}, $^{\rm d}$\citet{Boogert08}, $^{\rm e}$\citet{Bottinelli10}, $^{\rm f}$\citet{Oberg08}
\end{center}
}
\end{table*} 

Interstellar complex organic molecules or COMs\footnote{for the purpose of this paper COMs are hydrogen-rich organics containing at least three heavy elements, i.e. the kind of organics typically associated with hot cores} are the proposed starting point of an even more complex, prebiotic chemistry during star and planet formation, linking interstellar chemistry with the origins of life \citep{Herbst09}. Determining COM abundance patterns are important to constrain the reservoirs of organic material during the formation of stars and planetary systems, and to elucidate COM formation mechanisms. COMs were first detected in the hot cores associated with high-mass star formation \citep[e.g.]{Blake87,Helmich97}, but during the past decade COMs have been detected in an increasingly diverse set of environments, including pre-stellar cores, protostellar envelopes, outflows and hot cores in low-mass star forming regions \citep{Cazaux03,Bottinelli04b,Bottinelli07,Arce08,Oberg10c,Oberg11b,Bacmann12,Cernicharo12}. These detections suggest the existence of robust formation pathways of COMs, and hence that COMs might be common during the formation of low-mass or Solar-type stars. 

Based on the observed abundances of COMs and the diversity of their hosts, most complex molecules are proposed to form on the surfaces of interstellar dust grains and in icy grain mantles \citep{Herbst09}. Atom addition reactions on grains should be efficient at all temperatures, but may be mainly important to form smaller organics such as CH$_3$OH. Ice photodissociation followed by diffusion and radical-radical combination reactions in the ice should result in an efficient formation of complex molecules at slightly elevated temperatures (T$>$30~K) \citep{Garrod06, Herbst09}.  In this scenario, both the initial ice composition and the level of heating and UV processing should impact on the amount of complex molecules observed in a particular source. This scenario is supported by observations on that the initial ice composition is correlated with protostellar chemistry \citep{Oberg09a,Sakai10}, and claims of a different COM abundance pattern towards low-mass and high-mass YSOs \citep{Oberg11c,Caselli12}. These claims suffer from small-number-statistics, however; there are only $\sim$10 low-mass young stellar objects with reported detections of COMs, and the abundance distributions  of COMs during low-mass star formation is therefore poorly constrained. 

In this paper we present the result of a six-object pilot survey of COMs toward low-mass protostars using the IRAM 30m telescope. The sample is presented in \S1 and the observations and data reduction are described in \S2. In \S3 we present an overview of the spectral line data. We determine the column densities and, where possible, excitation temperatures of CH$_3$OH and detected COMs. We then use these new results together with existing literature detections and upper limits of representative COMs to obtain a first estimate of the frequency distribution of COMs toward low-mass YSOs. The results are compared with models and massive YSO chemistry in \S4.

\section{Source sample \label{sample}}

Our sources were selected from the $c2d$ (cores to disk) ice sample \citep{Boogert08}. The $c2d$ ice sample is a sub-sample of the $c2d$ survey of young stellar objects (YSOs) in nearby star-forming regions, i.e. the Perseus, Taurus, Serpens, and Corona Australis molecular cloud complexes, and a number of nearby isolated dense cores \citep{Evans03}. The spectral energy distributions of the ice sample span a range of IR spectral indices, $\alpha= (-0.25)-2.70$, where $\alpha$ is defined as the slope between 2 and 24 $\mu$m. In the infrared classification scheme $\alpha>0.3$ defines class 0/I sources \citep{Wilking01}, which are often, but not always, associated with young, embedded YSOs. 

\begin{figure*}%[htp]
\epsscale{1.0}
\plotone{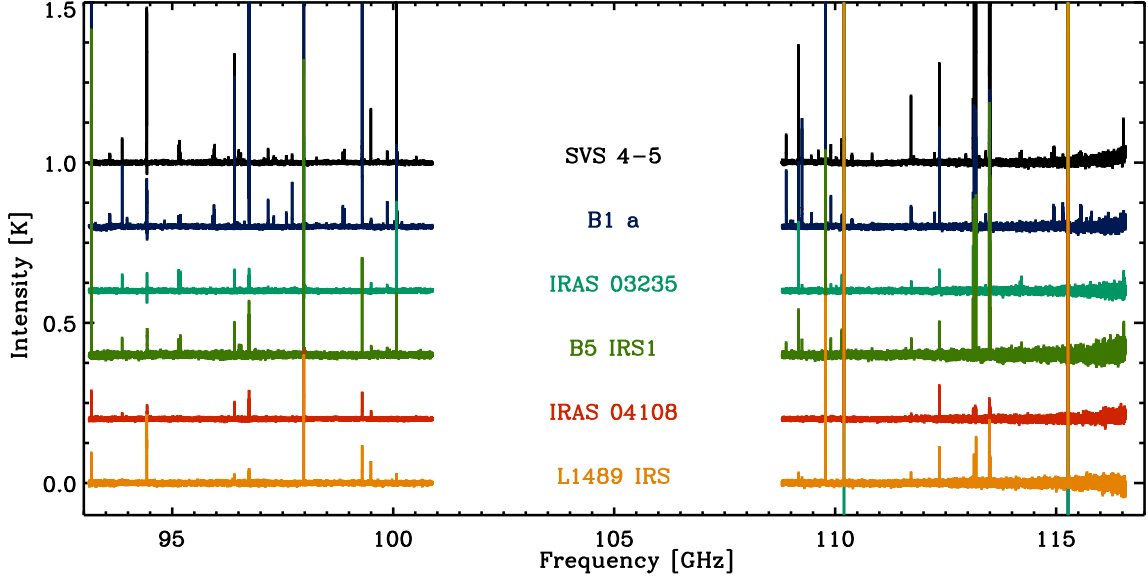}
\caption{93--101 and 109--117 GHz IRAM 30m spectra toward the LYSO sample, displaying the large variety in spectral line density between the different sources. Toward the top two sources, B1-a and SVS 4-5, many of the weak lines belong to complex organics. \label{fig:sample}}
\end{figure*}

From this sample we considered sources  on the northern sky with $\alpha>0.3$, a H$_2$O ice column $>2\times10^{18}$ cm$^{-2}$, i.e. young embedded low-mass protostars. From the 19 objects that fulfill these criteria in the $c2d$ ice sample, we selected six sources that sample the observed range of ice abundances, based on CH$_3$OH/NH$_3$ and CH$_3$OH/H$_2$O abundance ratios. Table \ref{sources} lists the source coordinates, bolometric luminosities, envelope masses and the IR SED indices, together with the ice abundances. B1-b, which was the target of a previous COM search, is also included in the table \citep{Oberg10c}. Together these seven sources span H$_2$O ice column densities between 2 and 14$\times10^{18}$, which can be compared with the complete $c2d$ ice sample, where $N_{\rm CH_3OH}=0.4-39\times10^{18}$ cm$^{-2}$. The sources come from three different clouds, Perseus, Taurus and Serpens, and where measured the envelopes are a few solar masses and the luminosities between one and 10s of solar luminosities. $N_{\rm CH_3OH}/N_{\rm H_2O}=2-25\%$, which is similar to the complete sample and so is range of ratios $N_{\rm CH_3OH}/N_{\rm NH_3}=0.5-6\%$.

None of the selected sources have been searched for COMs previously, but several have been detected in CH$_3$OH \citep{Oberg09a}, which is known to correlate with O-bearing COMs in hot cores  \citep{Bisschop07}. In summary this is a small sample, but it has been selected to be as representative as possible from the larger $c2d$ ice sample and should thus provide a first constraint on the prevalence of complex organics during the embedded phases of low-mass star formation.

\section{Observations \label{sec:obs}}

IRAS 03235+3004, B1-a, B5 IRS 1, L1489 IRS, IRAS 04108+2803 and SVS 4--5 were observed with the IRAM 30m Telescope on June 12--16, 2013  using the EMIR 90 GHz receiver and the FTS backend. The two sidebands cover 93--101~GHz and 109--117~GHz at a spectral resolution of 200~kHz or $\sim$0.5-0.6~km~s$^{-1}$ (Fig. \ref{fig:sample}) and with a sideband rejection of -15dB \citep{Carter12}. This spectral set-up was selected because of the potential large number of complex organic lines at these frequencies,  and the presence of the CH$_3$OH 2--1 ladder.

The pointing positions are listed in Table 1 and pointing was checked every 1--2~h and found to be accurate within 2--3$\arcsec$. Focus was checked every 4~h, and generally remained stable through most of the observations, i.e. corrections of $<$0.4~mm. Observations were acquired using both the position switching and wobbler switching modes. The position switching mode was attempted because of possible extended emission, but was found to have severe baseline instabilities. Comparison of the wobbler and position switch spectra revealed no significant absorption in the wobbler off-position in any of the sources, hence we only used the higher-quality wobbler spectra in this paper. The total integration time in the wobbler mode was $\sim$2--5~h for each source, under average to good summer weather conditions ($\tau=0.1-0.4$), resulting in a T$_{\rm a}^*$ rms of 3.5--7~mK in the lower sideband. B5 IRS1 and L1489 IRS were the only two sources with an rms above 5~mK. 

The spectra were reduced using CLASS\\ (http://www.iram.fr/IRAMFR/GILDAS). A global baseline was fit to each 4~GHz spectral chunk using 4--7 windows. The individual scans were baseline subtracted and averaged. To convert from antenna temperature, T$_{\rm a}^*$, to main beam temperature, T$_{\rm mb}$, forward efficiencies and beam efficiencies of 0.95 and 0.81 were applied. The spectra were converted to rest frequency using literature source velocities, fine-tuned using the frequencies of the CH$_3$OH 2--1 ladder. The absolute calibration of the spectra was also checked by comparing the CH$_3$OH 2--1 ladder with previous observations of some of the same sources, and were found to agree within 10\% \citep{Oberg09a}.

\section{Results \label{sec:res}}

\subsection{Spectral analysis}

Figure \ref{fig:sample} shows the complete 16 GHz spectra toward all six sources, ordered in terms of line richness. Two of the YSOs, SVS 4-5 and B1-a, stand out as particularly line rich. Both sources are in the vicinity of two other YSOs, SMM4 and B1-b, that are known hosts of complex molecules. Of the remaining four sources, B5 IRS1 and IRAS 03235+2004 are more line dense compared to the two Taurus sources L1489 IRS and IRAS 04108+2803. 

\begin{figure}%[htp]
\epsscale{1.0}
\plotone{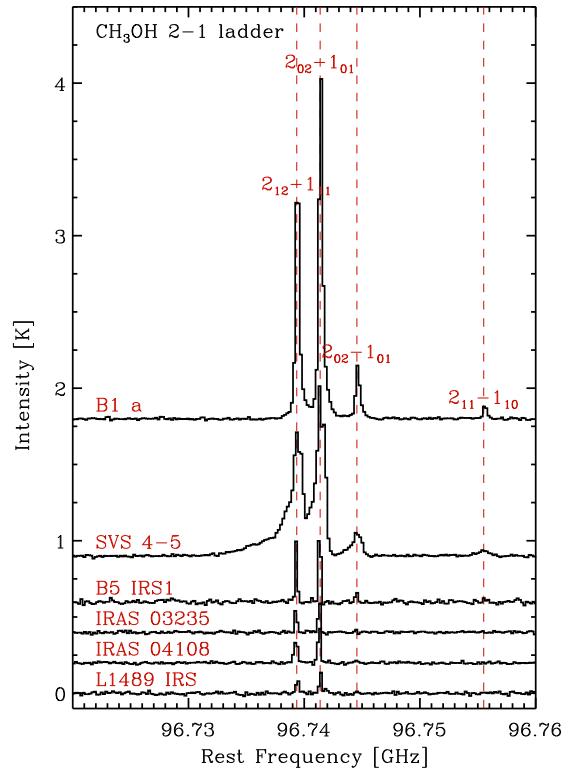}
\caption{CH$_3$OH 2--1 spectra toward the LYSO sample displaying the order of magnitude difference in line intensities across the sample. The spectra have been shifted with the systemic velocity of each source.  \label{fig:ch3oh}}
\end{figure}

These differences in line density seems correlated with the strength of the CH$_3$OH 2--1 ladder as shown in Fig. \ref{fig:ch3oh}. SVS 4-3 and B1-a both display strong CH$_3$OH lines, B5 IRS1 moderate ones, and IRAS 03235, IRAS 04108 and L1489 IRS very weak lines (peak intensities $<0.2$ K). In addition to the 2--1 lines there are a handful of other CH$_3$OH lines throughout the observed frequency range with excitation energies of 6--83~K (as well as higher energy lines which are not detected toward any source). Table \ref{tab:ch3oh} lists the integrated line intensities or 3$\sigma$ upper limits of all CH$_3$OH lines detected toward at least one source.  The integrated intensities were determined using IDL and MPFIT to fit Gaussians to the expected line positions. The 1$\sigma$ integrated line intensity uncertainties were extracted from the fit procedure and are often larger than the the 1$\sigma$ rms because it includes the fit uncertainty.  3$\sigma$ upper limits were determined using the rms in each 4 GHz chunk and the average CH$_3$OH line FWHM (Table \ref{fwhm}) for each source. Table \ref{tab:ch3oh} also presents the integrated line intensities of two $^{13}$CH$_3$OH lines detected toward B1-a and SVS 4-5.

\begin{table*}%[htp]
{\footnotesize
\begin{center}
\caption{Integrated CH$_3$OH line intensities in K km s$^{-1}$.\label{tab:ch3oh}}
\begin{tabular}{l c c c c c c c c}
\hline\hline
Freq & QN	&E$_{\rm up}$ &B1-a &SVS 4-5 &B5 IRS1 &IRAS 03235 &IRAS 04108 &L1489 IRS\\
\hline
$^{12}CH_3OH$\\
 95.169&  	$8_{08}-7_{17}$	&83& 0.119[0.012]& 0.297[0.012]& 0.030[0.009]& 0.016[0.011]& $<$ 0.013& $<$ 0.019\\
 95.914&  	$2_{12}-1_{11}	$	&21& 0.102[0.012]& 0.100[0.017]& 0.033[0.019]&$<$ 0.011& 0.017[0.012]& $<$ 0.019\\
 96.739&  $2_{12}-1_{11}	$	&12& 2.280[0.010]& 3.325[0.019]& 0.293[0.012]& 0.161[0.006]& 0.186[0.009]& 0.092[0.014]\\
 96.741&   $2_{02}-1_{01}$	&6& 3.113[0.009]& 3.272[0.015]& 0.514[0.009]&0.179[0.085]& 0.225[0.014]& 0.116[0.010]\\
 96.745&   $2_{02}-1_{01}$	&  20& 0.509[0.010]& 0.487[0.015]& 0.057[0.124]& $<$0.011& 0.017[0.008]& $<$0.019\\
 96.756&   $2_{11}-1_{10}$	&  28& 0.113[0.010]& 0.138[0.018]& $<$0.018&$<$0.011& $<$0.013& $<$0.019\\
 97.583&   $2_{11}-1_{10}$	&  21& 0.119[0.009]& 0.118[0.015]& $<$ 0.018&$<$0.011& $<$0.013& $<$0.019\\
108.894&   $0_{00}-1_{11}$	&  13& 0.437[0.011]& 0.429[0.018]& 0.080[0.022]&$<$ 0.012& $<$ 0.013 &$<$ 0.021\\
$^{13}CH_3OH$\\
94.405&	   $2_{12}-1_{11}$	&12	&0.040[0.009]	&0.047[0.012]	&$<$0.018	&$<$0.011	&$<$0.013	&$<$0.019\\
94.407&	   $2_{02}-1_{01}$	&7	&0.052[0.0012]	&0.043[0.013]	&$<$0.018	&$<$0.011	&$<$0.013	&$<$0.019\\
\hline
\end{tabular}
\end{center}
}
\end{table*} 

\begin{figure*}%[htp]
\epsscale{1.0}
\plotone{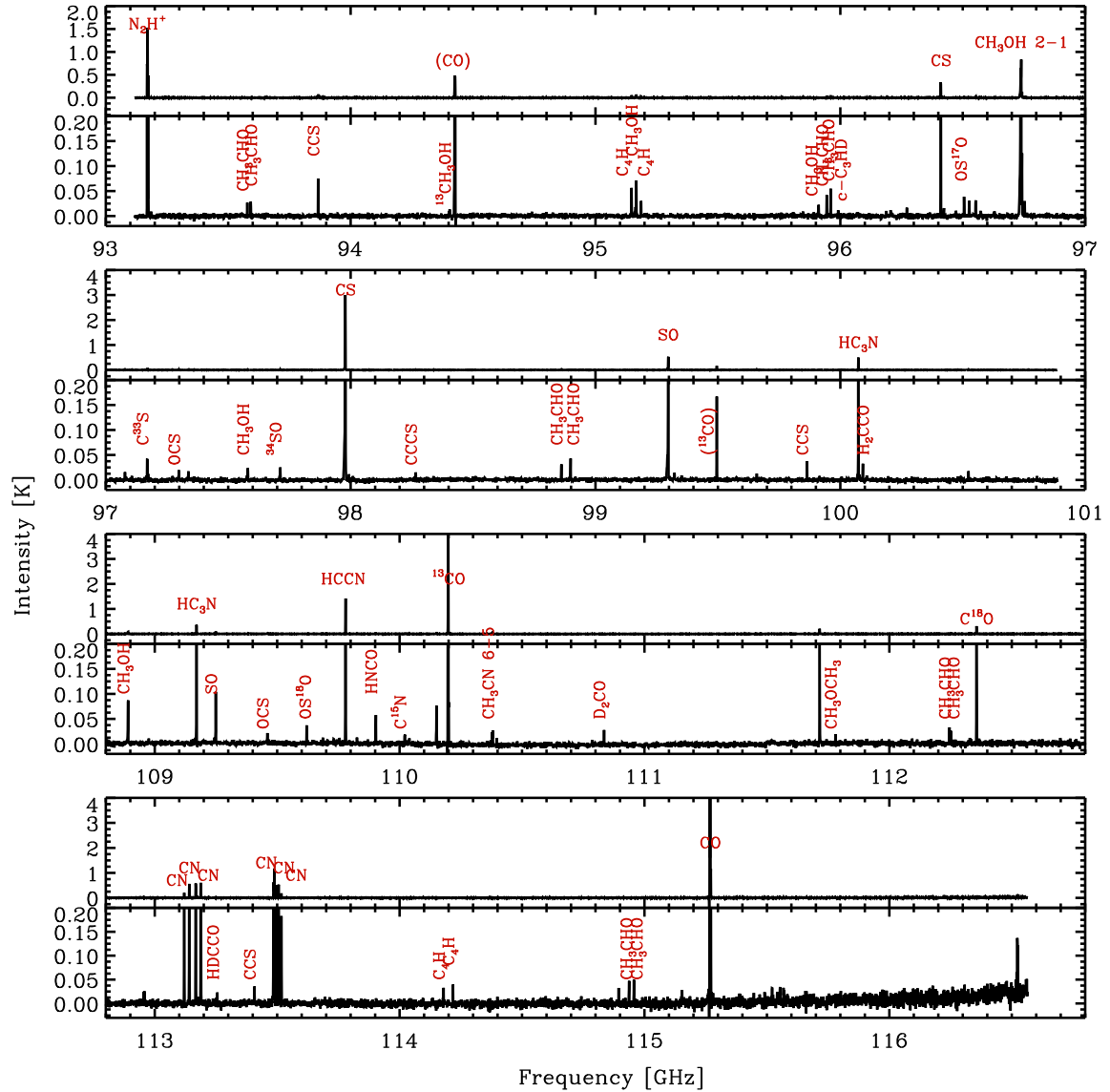}
\caption{Spectra at 93--101 and 109--117 GHz toward SVS 4--5 with key lines identified. For each 4~GHz frequency chunk a zoom-out and zoomed-in view is shown to visualize the range of line strengths, and the presence of both simple and more complex molecules in this frequency range. Notable identifications are CH$_3$OH, CH$_3$CHO, CH$_3$OCH$_3$, H$_2$CCO, HNCO and CH$_3$CN. Note that CO and $^{13}$CO both presents line `ghosts' in the lower sideband spectrum.  \label{fig:svs}}
\end{figure*}

Figure \ref{fig:svs} shows that in the case of SVS 4-5 (and B1-a) the multitude of weak lines hinted at in Figure \ref{fig:sample} are to a large degree associated with the complex organic molecules HNCO, H$_2$CCO, CH$_3$CHO, CH$_3$OCH$_3$ and CH$_3$CN. Line identifications were made using the Splatalogue web tool drawing upon the CDMS and JPL spectral databases \citep{Pickett98,Muller01}. Other lines are identified with simple molecules, CH$_3$OH and carbon chains (carbon chain abundances will be the topic of a future publication).

\begin{figure*}%[htp]
\epsscale{1.0}
\plotone{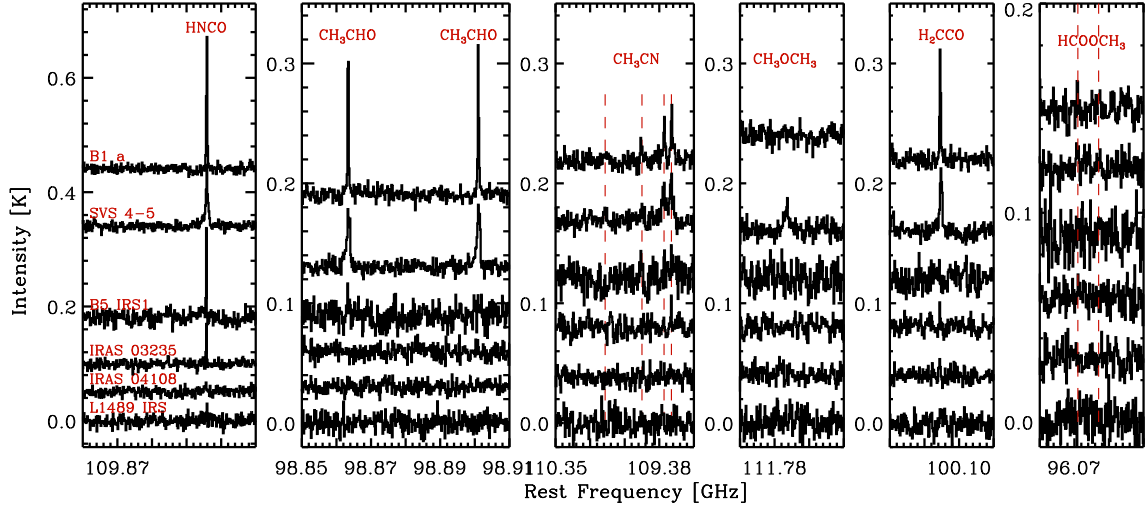}
\caption{Zoomed in spectra of HNCO, CH$_3$CHO, CH$_3$CN, CH$_3$OCH$_3$, H$_2$CCO and HCOOCH$_3$,  toward the LYSO sample. All spectra have been shifted by the source systemic velocities. \label{fig:com}}
\end{figure*}

Figure \ref{fig:com} shows blow-ups of the spectral regions with the strongest lines of the six complex organics detected toward at least one source. HNCO is detected toward all sources in the sample. CH$_3$CHO and CH$_3$CN  are clearly detected toward SVS 4-5 and B1-a, and marginally toward B5 IRS1. CH$_3$OCH$_3$ is only detected toward B1-a. H$_2$CCO is detected toward B1 a, SVS 4-5, IRAS 03235 and IRAS 04108. HCOOCH$_3$ is marginally detected toward SVS 4-5 and B1-a (the reality of these 3$\sigma$ detections are supported by several more marginal detections throughout the spectral range). In general we claim marginal detections for line intensities that exceed the 3$\sigma$ upper limit for that source and sideband. To count as a clear detection we furthermore require that the Gaussian fit is sufficiently well-defined to result in an integrated intensity estimate of 30\% or less. Table \ref{tab:coms} lists the detected COM line intensities and upper limits, calculated using Gaussian fits as described for CH$_3$OH lines above. 

\begin{table}%[htp]
{\footnotesize
\begin{center}
\caption{Line FWHM with standard deviations.\label{fwhm}}
\begin{tabular}{l c c c c c c c c}
\hline\hline
Source	&FWHM$_{\rm CH_3OH}$ \\%\\&FWHM$_{\rm HNCO}$&FWHM$_{\rm H_2CCO}$&FWHM$_{\rm CH_3CHO}$ &FWHM$_{\rm CH_3CN}$\\
& [km/s]\\%& [km/s]& [km/s]& [km/s]& [km/s]\\
\hline
\hline
B1-a 		&1.6[0.4]\\
SVS 4-5 		&3.7[0.7]\\
B5 IRS1 		&0.8[0.1]\\
IRAS 03235 	&0.9[0.5]\\
IRAS 04108 	&1.2[0.6]\\
L1489 IRS	&1.6[1.0]\\
\hline
\end{tabular}
\end{center}
}
\end{table} 

\begin{table*}%[htp]
{\footnotesize
\begin{center}
\caption{Integrated line intensities in K km s$^{-1}$ of complex organic molecules. \label{tab:coms}}
\begin{tabular}{l c c c c c c c c}
\hline\hline
Freq & E$_{\rm up}$  &B1-a &SVS 4-5 &B5 IRS1&IRAS 03235 &IRAS 04108 &L1489 IRS \\
\hline
{\it HNCO}\\
109.906&  15& 0.220[0.009]& 0.200[0.013]& 0.106[0.019]& 0.050[0.008]& 0.031[0.014]& 0.027[0.007]\\
{\it H$_2$CCO}\\
100.095&  27& 0.110[0.008]& 0.099[0.011]&  $<$0.011& 0.020[0.012]& 0.027[0.010]&  $<$0.017\\
{\it CH$_3$CN}\\
110.375&  47& 0.018[0.006]& 0.017[0.011]& 0.031[0.011]&$<$ 0.012& $<$ 0.013 &$<$ 0.02\\
110.381&  25& 0.046[0.011]& 0.073[0.015]& 0.024[0.008]&$<$ 0.012& $<$ 0.013 &$<$ 0.021\\
110.383&  18& 0.062[0.011]& 0.085[0.015]& 0.020[0.007]&$<$ 0.012& $<$ 0.013 &$<$ 0.021\\
{\it CH$_3$CHO}\\
93.581&  15& 0.098[0.008]& 0.115[0.015]& 0.016[0.008]&$<$ 0.011& $<$ 0.013 &$<$ 0.019\\
 93.595&  15& 0.106[0.010]& 0.125[0.014]& $<$0.018&$<$ 0.011& $<$ 0.013 &$<$ 0.019\\
 95.947&  13& 0.161[0.007]& 0.204[0.014]&  $<$0.018&$<$ 0.011& $<$ 0.013 &$<$ 0.019\\
 95.963&  13& 0.176[0.008]& 0.204[0.014]&  $<$0.018&$<$ 0.011& $<$ 0.013 &$<$ 0.019\\
96.274&  22& 0.042[0.009]& 0.061[0.014]&  $<$0.018&$<$ 0.011& $<$ 0.013 &$<$ 0.019\\
 96.426&  22& 0.033[0.009]& 0.065[0.013]&  $<$0.016&$<$ 0.010& $<$ 0.011 &$<$ 0.017\\
 96.476&  23& 0.043[0.010]& 0.045[0.014]&  $<$0.016&$<$ 0.010& $<$ 0.011 &$<$ 0.017\\
 96.633&  22& 0.050[0.009]& 0.039[0.013]& $<$0.016&$<$ 0.010& $<$ 0.011 &$<$ 0.017\\
 98.863&  16& 0.155[0.009]& 0.162[0.014]& 0.021[0.006]&$<$ 0.010& $<$ 0.011 &$<$ 0.017\\
 98.901&  16& 0.138[0.007]& 0.168[0.013]& 0.041[0.015]&$<$ 0.010& $<$ 0.011 &$<$ 0.017\\
112.249&  21& 0.096[0.010]& 0.131[0.016]& 0.015[0.008]&$<$ 0.012& $<$ 0.013 &$<$ 0.021\\
112.255&  21& 0.107[0.009]& 0.149[0.019]& 0.032[0.021]&$<$ 0.012& $<$ 0.013 &$<$ 0.021\\
{\it CH$_3$OCH$_3$}\\
111.783&  25& $<$0.013& 0.085[0.019]& $<$0.019&$<$ 0.012& $<$ 0.013&$<$ 0.021\\
115.545&  14& $<$0.027& 0.080[0.028]& $<$0.032&$<$ 0.019& $<$ 0.024 &$<$ 0.035\\
{\it HCOOCH$_3$}\\
 96.071&  23& 0.019[0.004]& 0.022[0.013]&$<$0.018& $<$ 0.010& $<$ 0.011 &$<$ 0.017\\
 96.077&  23& 0.018[0.011]& $<$0.032& $<$0.018 &$<$ 0.010& $<$ 0.011 &$<$ 0.017\\
100.482&  22& 0.014[0.011]& 0.023[0.010]&  $<$0.011& $<$ 0.010& $<$ 0.011 &$<$ 0.017\\
100.491&  22& 0.020[0.010]& 0.035[0.015]&  $<$0.011& $<$ 0.010& $<$ 0.011 &$<$ 0.017\\
100.683&  24& 0.018[0.009]& 0.044[0.013]&  $<$0.011& $<$ 0.010& $<$ 0.011 &$<$ 0.017\\
111.674&  28& 0.022[0.013]& 0.020[0.011]& $<$0.019&$<$ 0.012& $<$0.013&$<$ 0.021\\
111.682&  28& 0.019[0.017]& 0.010[0.011]& $<$0.019&$<$ 0.012&$<$0.013&$<$ 0.021\\
\hline
\end{tabular}
\end{center}
}
\end{table*}

\subsection{CH$_3$OH and COM abundances}

\begin{figure}%[htp]
\epsscale{1.0}
\plotone{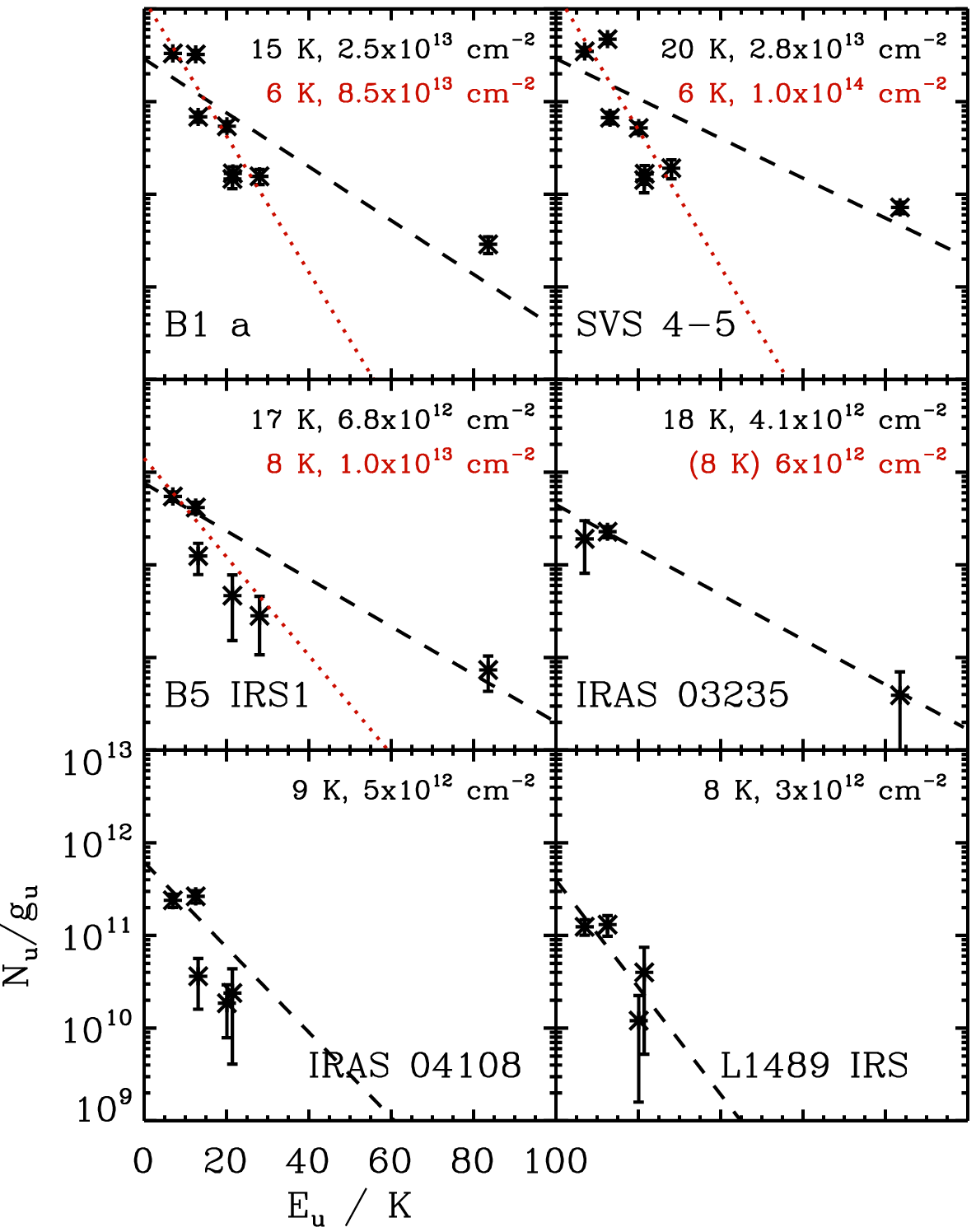}
\caption{CH$_3$OH rotational diagrams. Black lines and labels are for the fit to all lines. Red lines and labels are for fits that exclude high-energy lines. \label{fig:rot_dia}}
\end{figure}

The CH$_3$OH excitation temperatures and column densities were determined using the rotational diagram method \citep{Goldsmith99}, assuming optically thin lines and LTE at a single temperature -- the validity of these assumptions and the constraints on the kinetic temperatures provided by the excitation temperatures are explored further in \S\ref{lvg}. Figure \ref{fig:rot_dia} shows the results for CH$_3$OH. A single fit to all lines result in excitation temperatures of 15--20~K for the sources with any high energy lines detected. 

The observed CH$_3$OH emission does not generally seem to be well described by a single excitation temperature, however, but rather seems to trace a warm and cold component. Too few high energy lines are detected to quantify the warm component, but the cold component excitation temperatures and column densities can be estimated by focusing on the low-energy lines for the fit. This result in excitation temperatures of 6--8~K. The derived column densities should be representative of the protostellar envelopes outside of the core region where thermal evaporation is possible, i.e. at T$<$100~K -- because CH$_3$OH is readily sub-thermally excited it is not possible to {\it a priori} constrain the emission region any further. For IRAS 03235, too few low energy lines were detected to determine an envelope temperature, and an excitation temperature of 8~K was assumed to constrain the envelope column density.

To estimate the amount of warm CH$_3$OH column (beam-averaged) in each line of sight we use an excitation temperature of 26~K, based on the CH$_3$CN analysis below, and the one high-energy CH$_3$OH line excluded in the cold component fit. Table \ref{ch3oh_ab} reports the excitation temperatures and column densities calculated for the three fits characterizing the total, the cold and the warm protostellar CH$_3$OH column. The cold column densities range between $0.5-10\times10^{13}$ cm$^{-2}$, the warm column densities range between $<0.12-9\times10^{13}$ cm$^{-2}$ and the total column densities range between $0.5-23\times10^{13}$ cm$^{-2}$. The column density uncertainties includes a 10\% calibration uncertainty in addition to the fit uncertainty. The listed uncertainties do not incorporate the fact that the cold component will contribute slightly to the high energy line intensity and vice versa, resulting in systematic overestimates of the cold and warm component column densities, and indeed the sum of the cold+warm component column densities is 10--50\% higher compared to the column densities derived from the single component fit. The component column densities are thus at best accurate within a factor of 2.

\begin{table*}%[htp]
{\footnotesize
\begin{center}
\caption{CH$_3$OH excitation temperatures and column densities. The listed uncertainties are the formal errors; the systematic column density uncertainties are estimated to $\sim$50\% based on LVG analysis of B1-a. \label{ch3oh_ab}}
\begin{tabular}{l c c c c c c c c}
\hline\hline
Source	&T$_{\rm rot}$ ave & N$_{\rm CH_3OH}$ ave &T$_{\rm rot}$ cold & N$_{\rm CH_3OH}$ cold &T$_{\rm rot}$ warm$^{\rm b}$ & N$_{\rm CH_3OH}$ warm$^{\rm c}$ \\
	&[K] 	&[cm$^{-2}$]&[K] 	&[cm$^{-2}$]&[K] 	&[cm$^{-2}$]\\
\hline
B1-a 		&14.9[0.6]	&1.3[0.2]$\times10^{14}$	&5.9[0.3]	&8.4[1.9]$\times10^{13}$	&(26)	&9[3]$\times10^{13}$\\
SVS 4-5 		&20.2[0.9]	&2.3[0.4]$\times10^{14}$	&5.9[0.3]	&10[2]$\times10^{13}$	&(26)	&2.2[0.5]$\times10^{14}$\\
B5 IRS1 		&17[2]	&4.3[0.9]$\times10^{13}$	&8[1]		&2.1[0.7]$\times10^{13}$	&(26)	&2.3[1.2]$\times10^{13}$\\
IRAS 03235 	&18[4]	&2.8[0.9]$\times10^{13}$	&(8)$^{\rm a}$		&1.1[0.5]$\times10^{13}$	&(26)	&1.2[1.1]$\times10^{13}$\\
IRAS 04108 	&9[3]		&1.2[0.5]$\times10^{13}$	&9[3]		&1.2[0.5]$\times10^{13}$	&(26)	&$<8\times10^{12}$\\
L1489 IRS	&8[4]		&5[5]$\times10^{12}$	&8[4]		&5[5]$\times10^{12}$	&(26)	&$<1.3\times10^{12}$\\
\hline
\end{tabular}
\\$^{\rm a}$No excitation temperature could be independently derived and the sample average of 8~K was assumed.\\
$^{\rm b}$26 K is the assumed excitation temperature for the warm CH$_3$OH component.\\
$^{\rm c}$Based on the intensity or upper limit of the CH$_3$OH 95.169 GHz line.

\end{center}
}
\end{table*} 

\begin{table*}%[htp]
{\scriptsize
\begin{center}
\caption{Complex molecule excitation temperatures and column densities. \label{com_ab}}
\begin{tabular}{l c c c c c c c c c}
\hline\hline
Source	& \multicolumn{2}{c}{${\rm CH_3CHO}$} & \multicolumn{2}{c}{${\rm CH_3CN}$}& \multicolumn{2}{c}{${\rm HCOOCH_3}$} & ${\rm HNCO}^{\rm b}$ & ${\rm H_2CCO^{\rm b}}$ &${\rm CH_3OCH_3^{\rm b}}$\\
&T$_{\rm rot}^{\rm a}$ & N &T$_{\rm rot}^{\rm a}$ & N & T$_{\rm rot}^{\rm a}$ & N & N (T$_{\rm rot}=18\:K$) & N (T$_{\rm rot}=8\:K$)& N (T$_{\rm rot}=26\:K$)\\
&[K] &  [cm$^{-2}$]&[K] & [cm$^{-2}$]&[K] & [cm$^{-2}$]&  [cm$^{-2}$]& [cm$^{-2}$]&  [cm$^{-2}$]\\
\hline
\hline
B1-a 		&8[1]		&6[2]$\times10^{12}$	&26[8]	&1.2[0.6]$\times10^{12}$	&7[7]	&9[26]$\times10^{12}$	&2.6[0.3]$\times10^{12}$			&6.8[1.2]$\times10^{12}$		&$<6\times10^{12}$\\
SVS 4-5 		&9[2]		&7[2]$\times10^{12}$	&26[9]	&1.7[0.9]$\times10^{12}$	&8[9]	&1[3]$\times10^{13}$	&2.4[0.4]$\times10^{12}$			&5.9[1.3]$\times10^{12}$		&2.2[1.0]$\times10^{13}$\\
B5 IRS1 		&(8)		&1.9[0.7]$\times10^{12}$ 	&(26)	&4[2]$\times10^{11}$	&(8)	&$<4\times10^{12}$		&1.2[0.4]$\times10^{12}$			&$<9\times10^{11}$		&$<8\times10^{12}$\\
IRAS 03235 	&(8)		&$<7\times10^{11}$		&(26)	&$<3\times10^{11}$		&(8)	&$<2\times10^{12}$		&6.0[1.6]$\times10^{11}$			&1.2[0.8]$\times10^{12}$		&$<5\times10^{12}$\\
IRAS 04108 	&(8)		&$<6\times10^{11}$		&(26)	&$<4\times10^{11}$		&(8)	&$<1\times10^{12}$		&$4[2]\times10^{11}$			&1.6[0.9]$\times10^{12}$		&$8[4]\times10^{12}$\\
L1489 IRS	&(8)		&$<1.0\times10^{12}$	&(26)	&$<6\times10^{11}$		&(8)	&$<2\times10^{12}$		&3.2[1.1]$\times10^{11}$			&$2.5[1.1]\times10^{12}$	&$<8\times10^{12}$\\
\hline
\end{tabular}
\\$^{\rm a}$(T) is used to indicate that the sample average excitation temperature was used to derived a column density or upper limit.  \\
$^{\rm b}$ The average CH$_3$OH, CH$_3$CHO  and CH$_3$CN excitation temperatures were used to derive the HNCO, H$_2$CCO and CH$_3$OCH$_3$ column densities, respectively.
\end{center}
}
\end{table*} 

\begin{figure}%[htp]
\epsscale{1.}
\plotone{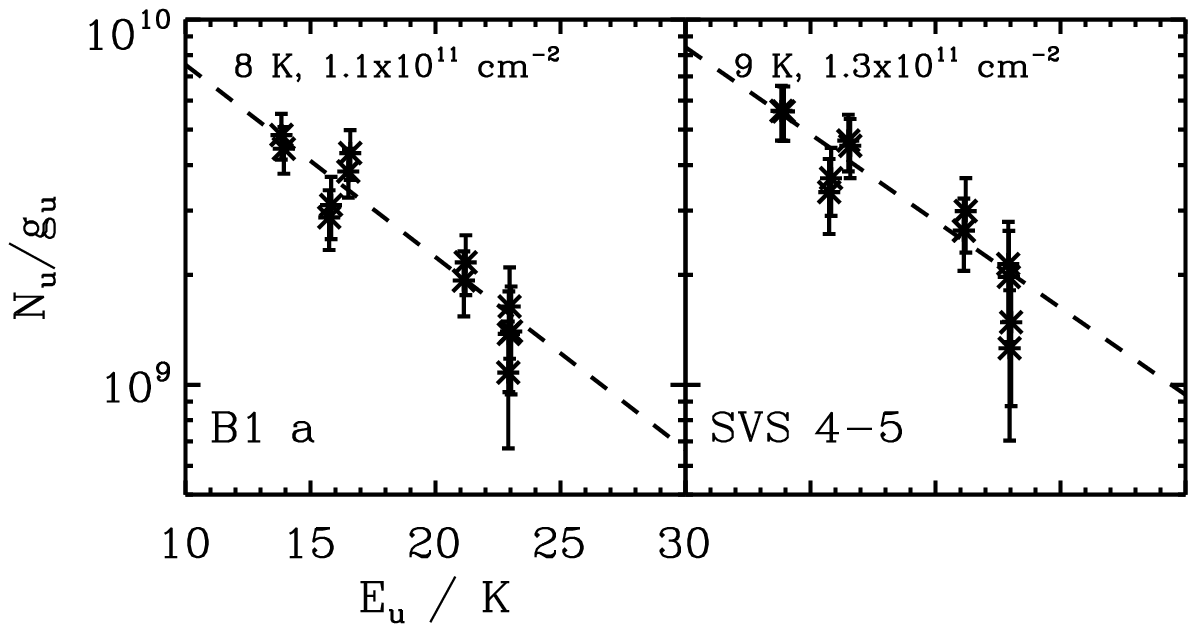}
\plotone{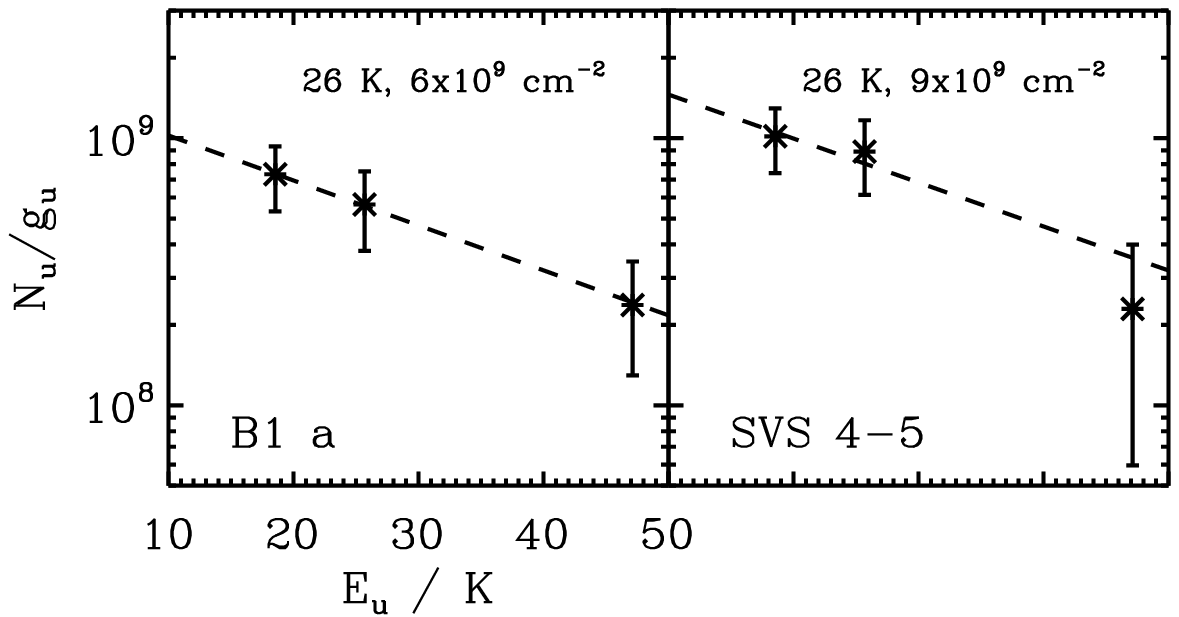}
\plotone{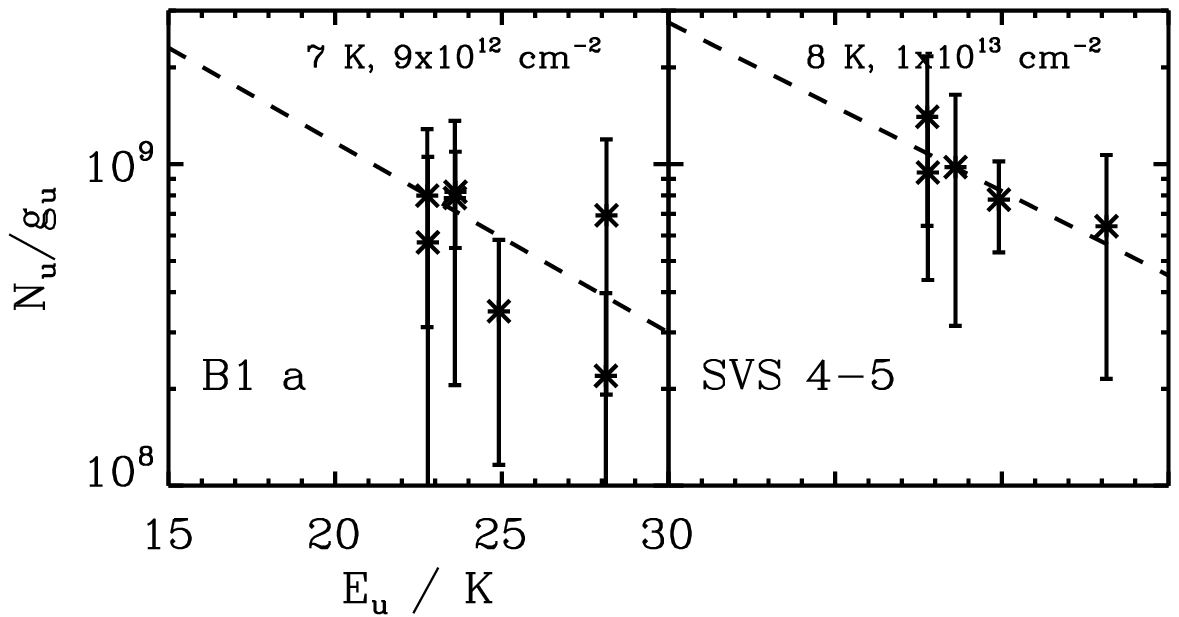}
\caption{CH$_3$CHO  (top), CH$_3$CN (middle), and HCOOCH$_3$ (bottom) rotational diagrams. \label{fig:rot_dia_hcooch3}}
\end{figure}

Figure \ref{fig:rot_dia_hcooch3} shows that CH$_3$CHO and CH$_3$CN line intensities can be fit by a single excitation temperature. Based on this analysis, CH$_3$CHO is characterized by a low excitation temperature (8--9~K) consistent with the CH$_3$OH cold component. This does not rule out the existence of a warm CH$_3$CHO component, since upper limits of higher energy lines are inconclusive, but the detected CH$_3$CHO does not originate in a hot core. In contrast the CH$_3$CN excitation temperature of 26~K is suggestive of a warm, possibly a hot core, origin for CH$_3$CN. The HCOOCH$_3$ detections have low SNR and also a small spread in energy levels and the derived excitation temperatures of 7--8~K are therefore highly uncertain. For CH$_3$OH as well as these COMs, the derived excitation temperatures depend on the kinetic temperatures. The two temperatures are rarely identical, however, and the relationship between them depend on the details of the excitation conditions and molecular excitation properties. In the case of CH$_3$CHO, CH$_3$CN and CH$_3$OH, comparable dipole moments, result in that the relative excitation temperatures constrain the relative excitation conditions of the different molecules, i.e. CH$_3$CN is more centrally peaked compared to CH$_3$CHO ({\it cf.} \S\ref{lvg}), even though the specific emission locations of the different molecules are unconstrained.

Table \ref{com_ab} lists the resulting excitation temperatures and beam-averaged column densities, i.e. the reported column densities do not take into account potential beam dilution, which may be substantial for some or all molecules. For the detected complex organic molecules, where it is not possible to determine an excitation temperature we picked a temperature, 8, 18 or 26~K, based on recent observations of spatial and excitation temperature correlations of different molecules around high-mass protostars \citep[][Fayolle et al. subm. ApJ]{Bisschop07,Oberg13}. In these studies H$_2$CCO and CH$_3$CHO are cold and we therefore used the CH$_3$CHO excitation temperature of 8~K to determine the H$_2$CCO column. CH$_3$OCH$_3$ and CH$_3$CN emission were generally associated with only warm material and we used the CH$_3$CN excitation temperature of 26~K to determine the CH$_3$OCH$_3$ column. HNCO, like CH$_3$OH, has been observed to trace both cold and warm material and we therefore used an excitation temperature of 18~K, typical for the CH$_3$OH when fitting a single LTE component to the protostellar data.  It is important to note that the excitation temperature choices are based on a small set of high-mass protostellar data and may need to be revised once more spatially resolved observations of complex molecules toward low-mass protostars exist. 

We used the derived complex molecule column densities together with the derived CH$_3$OH column densities to calculate the abundances of organic molecules with respect to CH$_3$OH in our source sample. We tested the sensitivity of the derived abundances on the selected CH$_3$OH components for different molecules, and generally found that the choice of excitation temperature and CH$_3$OH reference column density (beam-averaged, cold or warm components) affected the derived abundances with respect to CH$_3$OH by less than a factor of 2. Table \ref{abund} shows that the CH$_3$CHO abundances are between 7 and 9\% and the upper limits between 5 and 20\% with respect to CH$_3$OH. CH$_3$CN and CH$_3$OCH$_3$ abundances or upper limits could only be derived toward the first four sources since no lines associated with warm CH$_3$OH were detected toward two of the sources. The CH$_3$CN abundances are 0.8--1.7\% and the CH$_3$OCH$_3$ abundance toward SVS 4-5 is 10\%. The HCOOCH$_3$ abundances are $\sim$10\%, but abundances up to 40\% cannot be excluded. The HNCO abundances vary between 1.0--6.4\%, and the H$_2$CCO abundances vary between 1.1--9\%.  Derived upper limits are typically similar or higher compared to detections. 

\begin{table*}%[htp]
{\footnotesize
\begin{center}
\caption{Complex molecule abundances with respect to CH$_3$OH in \%. \label{abund}}
\begin{tabular}{l c c c c c c c c}
\hline\hline
Source	&x$_{\rm CH_3CHO}$ &x$_{\rm CH_3CN}$ & x$_{\rm HNCO}$ & x$_{\rm H_2CCO}$ & x$_{\rm CH_3OCH_3}$& x$_{\rm HCOOCH_3}$\\
\hline
B1-a 		&7[3]		&1.3[0.8]	&2.0[0.4]	&1.4[0.3]	&$<$7	&11[19]\\
SVS 4-5 		&7[2]		&0.8[0.4]		&1.0[0.3]	&1.1[0.2]	&10[5]	&10[30]\\
B5 IRS1 		&9[4]		&1.7[1.3]	&2.8[1.1]	&$<$0.8	&$<$35	&$<$21\\
IRAS 03235 	&$<$6	&$<$2.5	&2.1[0.9]	&2.0[1.5]	&$<$42	&$<$15\\
IRAS 04108 	&$<$5	&--$^{\rm a}$		&3.3[2.2]	&2.4[1.4]	&--$^{\rm a}$		&$<$12\\
L1489 IRS	&$<$20	&--$^{\rm a}$		&6.4[6.7]	&9[9]		&--$^{\rm a}$		&$<$46\\
\hline
\end{tabular}
\\$^{\rm a}$No upper limit could be derived because of lack of detection of warm CH$_3$OH.
\end{center}
}
\end{table*} 

\subsection{Optical depth and LVG modeling \label{lvg}}

\begin{figure}%[htp]
\epsscale{1.}
\plotone{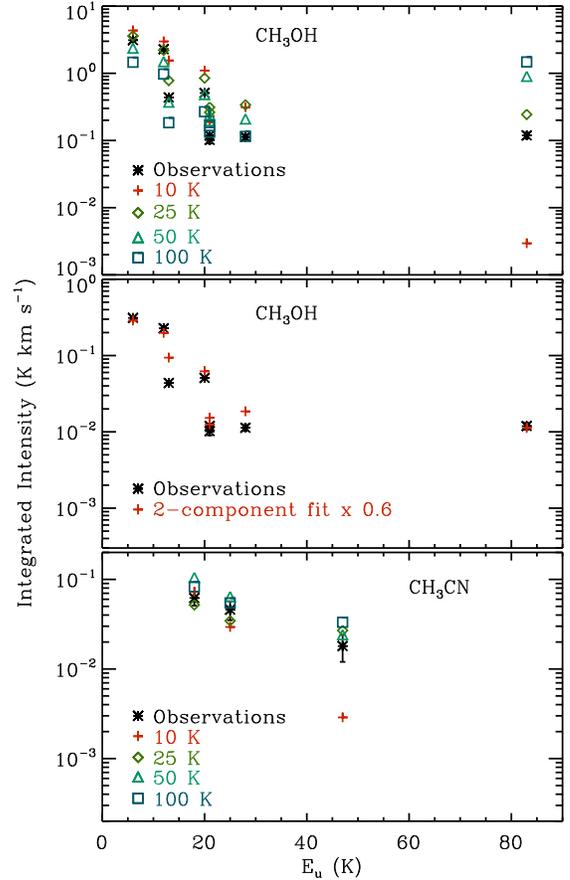}
\caption{Comparisons between observed and modeled line intensities toward B1-a using RADEX. The top panel show CH$_3$OH lines and models assuming a single component with constant temperature, a density of 10$^6$ cm$^{-3}$, and the derived average column from the rotational diagram analysis. The middle panel compares the same data with a two component fit, using the temperatures of the cold and warm component in Table \ref{ch3oh_ab}, but multiplying the column densities by 0.6. The bottom panel compares the observed and model CH$_3$CN line intensities. \label{fig:lvg}}
\end{figure}

The analysis in the previous section assumes that all observed complex organic emission is optically thin. This is a tenuous assumption for CH$_3$OH, since CH$_3$OH lines are known to sometimes be optically thick toward protostars \citep[e.g.][]{Bisschop07}. Fortunately, the $^{13}$CH$_3$OH 2--1 ladder falls within the observed frequency range, enabling a direct test of this assumption. Two $^{13}$CH$_3$OH lines are detected toward B1-a and SVS 4-5, but not toward any other sources. The line intensity ratios between corresponding CH$_3$OH and $^{13}$CH$_3$OH lines are 57--60[15] and 71--76[19] toward B1-a and SVS 4--5 $\sim$, respectively, consistent with the Solar $^{12}$C/$^{13}$C ratio of 77. CH$_3$OH, and by inference all other COM, emission lines thus appear to be optically thin.

A second question is how the excitation temperatures, derived using rotational diagrams, relate to the kinetic temperatures in the emission regions. To explore this, we carried out a Large Velocity Gradient (LVG) radiative transfer analysis using RADEX \citep{vanderTak07} for the molecules with both sufficient number of lines to constrain the excitation conditions and known collision cross sections, i.e. CH$_3$OH and CH$_3$CN, toward B1-a. B1-a is here taken to be representative of the sample because of the similar emission patterns of CH$_3$OH and  CH$_3$CN (when detected) across the sample (Figs. \ref{fig:rot_dia}--\ref{fig:rot_dia_hcooch3}). 

We ran a grid of RADEX models with $n_{\rm H}=[10^5,10^6]$ cm$^{-3}$, T$_{\rm kin}=[10,25,50,100]$~K, and CH$_3$OH and CH$_3$CN column densities ranging between 0.5$\times$ and 2$\times$ those listed in Tables \ref{tab:ch3oh}--\ref{tab:coms} for B1-a. The models with a factor of 2 lower and higher column densities compared to the rotational diagram values cannot be excluded for all temperatures and densities, but generally fit the data less well. This indicates that the rotational diagram method provides a good estimate of the beam-averaged column densities, but that the real uncertainty in the derived numbers are at least a factor of two rather than the formal uncertainties of 20--50\%.

Figure \ref{fig:lvg} shows the model results for the best fit column densities (i.e. the ones derived using the rotational diagram method), and a density of 10$^6$ cm$^{-3}$. The best fit temperature to the CH$_3$OH data, assuming a single component, is between 10 and 25~K, consistent with the rotational diagram excitation temperature. If a lower density of 10$^5$ cm$^{-2}$ is assumed, kinetic temperatures up to 50~K fit the data. However, such a low average density is unlikely based on \citet{Jorgensen02}, where a detailed envelope modeling shows that the average density toward deeply embedded protostars on 1000~AU scales is $\sim$10$^6$ cm$^{-3}$. In either case, it can be excluded that most of the CH$_3$OH emission originates in a hot core. As in the rotational diagram analysis, no single temperature component fits all the CH$_3$OH lines. The two components derived in Table \ref{ch3oh_ab} reproduces the relative line intensities very well, but over-predicts the line intensities by about 50\% as discussed above. In the case of CH$_3$CN, only kinetic temperatures of 25~K and higher are consistent with the data, confirming the rotational diagram results that CH$_3$CN has of a more centrally peaked origin compared to CH$_3$OH.

In summary, the analyzed emission lines are optically thin and the rotational diagram method provides an accurate (within a factor of 2) derivation of the molecular column densities. The excitation temperatures and the kinetic temperatures are not identical, but the observed difference in excitation temperatures for CH$_3$CN and CH$_3$OH corresponds to a real difference in kinetic temperatures for reasonable density assumptions.

\subsection{Sample statistics}

\begin{figure}%[htp]
\epsscale{1.0}
\plotone{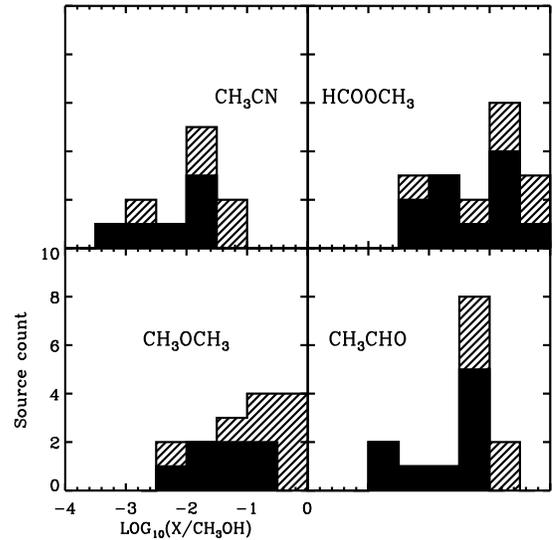}
\caption{Histograms of the abundances of selected complex organics with respect to CH$_3$OH taken from our sample and the literature. The solid bars are detections and line-filled bars indicated upper limits. \label{fig:stat}}
\end{figure}

In addition to the low-mass young stellar objects presented here, CH$_3$CHO, HCOOCH$_3$, CH$_3$CN, and/or CH$_3$OCH$_3$ have been previously quantified toward B1-b in Perseus \citep{Oberg09a,Oberg10a}, SMM1, SMM4 and SMM4-W in Serpens \citep{Oberg11b}, L1157 \citep{Arce08}, NGC 1333 IRAS 2A,  4A and 4B in Perseus \citep{Bottinelli04a,Bottinelli07}, and toward IRAS 16293-2422, both including the binary in one beam, and toward the A and B cores separately \citep{Cazaux03,Bisschop08}. The abundance frequency histograms with respect to CH$_3$OH are shown in Fig. \ref{fig:stat} for the combined low-mass sample of literature sources and our six YSOs. Complex molecule abundances above 1\% are clearly common. The median detected abundances for the four species HCOOCH$_3$, CH$_3$CHO, CH$_3$OCH$_3$ and CH$_3$CN varies between 1\% (CH$_3$CN) and 5\% (HCOOCH$_3$ and CH$_3$OCH$_3$) with respect to CH$_3$OH. It is important to note that most upper limits seems to be consistent with these values and thus that the lack of complex molecules in half of our sources is most likely due to an overall low CH$_3$OH abundance or a low overall ice desorption rate rather than a significantly different chemistry with respect to CH$_3$OH.

Figure \ref{fig:stat} also shows that the COM abundances vary significantly between sources. Disregarding upper limits CH$_3$CN/CH$_3$OH = 0.07--1.7\%, HCOOCH$_3$/CH$_3$OH = 0.6--56\%, CH$_3$OCH$_3$/CH$_3$OH = 0.8--20\% and CH$_3$CHO/CH$_3$OH = 0.1--9\%. For all molecules there is thus a one or two order of magnitude variation across the sample.

\section{Discussion \label{sec:disc}}

\begin{table}%[htp]
{\footnotesize
\begin{center}
\caption{Median COM abundances in LYSOs and massive YSOs with and without hot cores \label{comp}}
\begin{tabular}{l c c c c c c c c}
\hline\hline
Abundance \%CH$_3$OH	&LYSOs&Hot Cores$^{\rm a}$&MYSOs$^{\rm b}$\\
\hline
CH$_3$CN		&1	&7	&3\\
HCOOCH$_3$		&5	&12	&--\\
CH$_3$OCH$_3$	&5	&16	&4\\
CH$_3$CHO		&3	&0.1	&3\\
\hline
\end{tabular}
\\$^{\rm a}$\citet{Bisschop07}, $^{\rm b}$MYSOs without bright Hot Cores from Fayolle et al. (subm)
\end{center}
}
\end{table} 

The median abundances derived from the combined low-mass YSO sample can be compared both with the more well-studied high-mass YSO chemistry and with model predictions. In a sample of seven hot cores observed by \citet{Bisschop07} the median abundances of CH$_3$CN, HCOOCH$_3$, CH$_3$OCH$_3$ and CH$_3$CHO with respect to CH$_3$OH were 7\%, 12\%, 16\% and 0.1\%, respectively. That is, all abundances except for CH$_3$CHO are a factor of a few higher toward the high-mass hot cores. This may not be representative of massive YSO chemistry; in a recent sample of three high-mass YSOs without hot cores, the median abundances of CH$_3$CN , CH$_3$OCH$_3$ and CH$_3$CHO with respect to CH$_3$OH were all 3--4\% (Fayolle et al. ApJ subm.). These abundances are quite similar to our low-mass sample. There is thus no clear observational evidence for an intrinsic difference between low-mass and high-mass YSO COM chemistry with respect to CH$_3$OH, when averaged over all early stages of star formation. This does not exclude that there are significant differences when comparing low-mass and high-mass objects that are at the same evolutionary stage, i.e. there may still be real differences between high-mass hot cores and their low-mass equivalents because of e.g. different protostellar collapse time scales \citep{Garrod08}. More spatially resolved studies of low-mass YSOs are required, however, to support or reject that proposition.

This study also shows that there are orders of magnitude differences in COM abundances with respect to CH$_3$OH among low-mass YSOs. Three potential sources of variation are the initial ice composition, temperature and density structures and thus excitation conditions, and the chemical evolutionary stage. %The (lack of) observational constraints on their relative importance for the COM chemistry is evaluated qualitatively in the next few paragraphs -- the sample is too small for a quantitative assessment. 
A unique feature of this sample is that the ice compositions are known and its relationship to the Com chemistry can thus be evaluated. There is some correlation between CH$_3$OH ice and gas abundances among the 7 ice sources in Table 2, but there are also clear exceptions to this relationship, e.g. B1-a, which has a small CH$_3$OH ice abundance and a fairly high gas-phase abundance with respect to H$_2$O ice. There are no obvious correlations between the ice compositions and COM abundance patterns, in particular there is no correlation between NH$_3$/CH$_3$OH in the ice and N- and O-bearing complex molecules in the gas. This does not rule out that the ice composition is important for the complex chemistry, but in this small sample, ice composition is clearly not the dominant regulator of the observed COM chemistry.

In terms of evolutionary stage, the sample can be coarsely divided into ice sources (our sample + B1 b), embedded protostars without significant ice absorption (SMM 1, 4, and 4-W, NGC 1333 IRAS 2A, 4A and 4B, and IRAS 16293-2422), and an outflow (L1157). B1-b should potentially be in its own grouping since the COM emission seems more associated with the pre-stellar core than with the protostar. There is no significant difference between the ice sources and ice-free protostars in terms of COM emission, i.e. the median abundances within our sample are consistent with those of the ice-free protostars. The difference in envelope structure and/or evolutionary stage between these two source types thus also fails to explain the observed COM abundance variation across the sample.

Finally, assessing the importance of different excitation conditions is complicated by sparse information on the spatial origins of the complex molecule emission in the different sources. In the literature one out of two simplified distributions of COMs toward protostars is typically assumed: either that the CH$_3$OH and COM spatial distributions are the same and characterized by a single CH$_3$OH excitation temperature \citep[e.g.][]{Oberg11c}, or that the COM spatial distribution is more compact compared to CH$_3$OH \citep[e.g.][]{Bottinelli07}. In the latter case COM/CH$_3$OH abundances are calculated based on modeled CH$_3$OH core abundances and the assumption that all COM emission originates in the core. In this study we use the CH$_3$OH data to estimate the amount of cold CH$_3$OH in the envelope and warm CH$_3$OH in the core and then calculate COM/CH$_3$OH abundances using the cold or warm CH$_3$OH component or the total beam-averaged column density dependent on COM excitation temperatures and/or spatial distribution constraints from previous studies. This is a crude approximation and until spatially resolved observations of both CH$_3$OH and COMs exists toward a sample of low-mass YSOs, the reported abundances are estimated to be accurate only within a factor of a few, i.e. considerably less accurate than the $\sim$30--50\% uncertainties reported in Table \ref{abund} when only taking into account fit and calibration errors. There is clearly a need both to increase the existing number of sources searched for complex organics and to constrain the spatial distribution of COMs toward a sub-sample of representative sources to elucidate what source characteristics that set the observed COM abundances.

\section{Conclusions \label{sec:conc}}

We have carried out a small, pilot survey of complex molecules in a sample of low-mass YSOs, which were selected based on their measured ice abundances in the envelope. The results of this survey have been combined with literature values on complex molecule detections and upper limits to obtain first constraints on the abundance median and variability of complex molecule abundances during low-mass star formation. Based on this we have found:

\begin{enumerate}
\item Complex organics (CH$_3$CHO, HCOOCH$_3$, CH$_3$OCH$_3$ and/or CH$_3$CN) are detected toward 2--3/6 embedded protostars at abundances of 0.8--11\% with respect to CH$_3$OH (i.e. COMs are clearly detected toward two sources and marginally toward a third). Upper limits in the remaining sources are consistent with the detected abundances with respect to CH$_3$OH, indicative that complex molecule formation at the 1--10\% level with respect to CH$_3$OH is common during the early stages of low-mass star formation. 
\item Two other slightly less complex organic molecules, HNCO and H$_2$CCO are more common, and are clearly detected in 4--6 sources in the sample.
\item When the pilot survey is combined with 8 deep searches for complex organics in the literature we obtain median values with respect to CH$_3$OH of 1\% for CH$_3$CN, 3\% for CH$_3$CHO, 5\% for CH$_3$OCH$_3$, and 5\% for HCOOCH$_3$ for low-mass YSOs. 
\item There is at least an order of magnitude variability in abundances with respect to CH$_3$OH for all species, but the current sample is too small and heterogenous to constrain the origin of this variability to initial conditions, chemical evolution, or physical structures, or a combination of all three. 
\end{enumerate}

\acknowledgements
\noindent {\it Acknowledgements:} We gratefully acknowledge the IRAM staff for help provided during the observations and data reduction. 

\bibliographystyle{aa}

\begin{thebibliography}{36}
\expandafter\ifx\csname natexlab\endcsname\relax\def\natexlab#1{#1}\fi

\bibitem[{{Arce} {et~al.}(2008){Arce}, {Santiago-Garc{\'{\i}}a},
  {J{\o}rgensen}, {Tafalla}, \& {Bachiller}}]{Arce08}
{Arce}, H.~G., {Santiago-Garc{\'{\i}}a}, J., {J{\o}rgensen}, J.~K., {Tafalla},
  M., \& {Bachiller}, R. 2008, \apjl, 681, L21

\bibitem[{{Bacmann} {et~al.}(2012){Bacmann}, {Taquet}, {Faure}, {Kahane}, \&
  {Ceccarelli}}]{Bacmann12}
{Bacmann}, A., {Taquet}, V., {Faure}, A., {Kahane}, C., \& {Ceccarelli}, C.
  2012, \aap, 541, L12

\bibitem[{{Bisschop} {et~al.}(2008){Bisschop}, {J{\o}rgensen}, {Bourke},
  {Bottinelli}, \& {van Dishoeck}}]{Bisschop08}
{Bisschop}, S.~E., {J{\o}rgensen}, J.~K., {Bourke}, T.~L., {Bottinelli}, S., \&
  {van Dishoeck}, E.~F. 2008, \aap, 488, 959

\bibitem[{{Bisschop} {et~al.}(2007){Bisschop}, {J{\o}rgensen}, {van Dishoeck},
  \& {de Wachter}}]{Bisschop07}
{Bisschop}, S.~E., {J{\o}rgensen}, J.~K., {van Dishoeck}, E.~F., \& {de
  Wachter}, E.~B.~M. 2007, \aap, 465, 913

\bibitem[{{Blake} {et~al.}(1987){Blake}, {Sutton}, {Masson}, \&
  {Phillips}}]{Blake87}
{Blake}, G.~A., {Sutton}, E.~C., {Masson}, C.~R., \& {Phillips}, T.~G. 1987,
  \apj, 315, 621

\bibitem[{{Boogert} {et~al.}(2008){Boogert}, {Pontoppidan}, {Knez}, {Lahuis},
  {Kessler-Silacci}, {van Dishoeck}, {Blake}, {Augereau}, {Bisschop},
  {Bottinelli}, {Brooke}, {Brown}, {Crapsi}, {Evans}, {Fraser}, {Geers},
  {Huard}, {J{\o}rgensen}, {{\"O}berg}, {Allen}, {Harvey}, {Koerner}, {Mundy},
  {Padgett}, {Sargent}, \& {Stapelfeldt}}]{Boogert08}
{Boogert}, A.~C.~A., {Pontoppidan}, K.~M., {Knez}, C., {et~al.} 2008, \apj,
  678, 985

\bibitem[{{Bottinelli} {et~al.}(2010){Bottinelli}, {Boogert}, {Bouwman},
  {Beckwith}, {van Dishoeck}, {{\"O}berg}, {Pontoppidan}, {Linnartz}, {Blake},
  {Evans}, \& {Lahuis}}]{Bottinelli10}
{Bottinelli}, S., {Boogert}, A.~C.~A., {Bouwman}, J., {et~al.} 2010, \apj, 718,
  1100

\bibitem[{{Bottinelli} {et~al.}(2004{\natexlab{a}}){Bottinelli}, {Ceccarelli},
  {Lefloch}, {Williams}, {Castets}, {Caux}, {Cazaux}, {Maret}, {Parise}, \&
  {Tielens}}]{Bottinelli04a}
{Bottinelli}, S., {Ceccarelli}, C., {Lefloch}, B., {et~al.} 2004{\natexlab{a}},
  \apj, 615, 354

\bibitem[{{Bottinelli} {et~al.}(2004{\natexlab{b}}){Bottinelli}, {Ceccarelli},
  {Neri}, {Williams}, {Caux}, {Cazaux}, {Lefloch}, {Maret}, \&
  {Tielens}}]{Bottinelli04b}
{Bottinelli}, S., {Ceccarelli}, C., {Neri}, R., {et~al.} 2004{\natexlab{b}},
  \apjl, 617, L69

\bibitem[{{Bottinelli} {et~al.}(2007){Bottinelli}, {Ceccarelli}, {Williams}, \&
  {Lefloch}}]{Bottinelli07}
{Bottinelli}, S., {Ceccarelli}, C., {Williams}, J.~P., \& {Lefloch}, B. 2007,
  \aap, 463, 601

\bibitem[{{Carter} {et~al.}(2012){Carter}, {Lazareff}, {Maier}, {Chenu},
  {Fontana}, {Bortolotti}, {Boucher}, {Navarrini}, {Blanchet}, {Greve}, {John},
  {Kramer}, {Morel}, {Navarro}, {Pe{\~n}alver}, {Schuster}, \&
  {Thum}}]{Carter12}
{Carter}, M., {Lazareff}, B., {Maier}, D., {et~al.} 2012, \aap, 538, A89

\bibitem[{{Caselli} \& {Ceccarelli}(2012)}]{Caselli12}
{Caselli}, P. \& {Ceccarelli}, C. 2012, \aapr, 20, 56

\bibitem[{{Cazaux} {et~al.}(2003){Cazaux}, {Tielens}, {Ceccarelli}, {Castets},
  {Wakelam}, {Caux}, {Parise}, \& {Teyssier}}]{Cazaux03}
{Cazaux}, S., {Tielens}, A.~G.~G.~M., {Ceccarelli}, C., {et~al.} 2003, \apjl,
  593, L51

\bibitem[{{Cernicharo} {et~al.}(2012){Cernicharo}, {Marcelino}, {Roueff},
  {Gerin}, {Jim{\'e}nez-Escobar}, \& {Mu{\~n}oz Caro}}]{Cernicharo12}
{Cernicharo}, J., {Marcelino}, N., {Roueff}, E., {et~al.} 2012, \apjl, 759, L43

\bibitem[{{Evans} {et~al.}(2003){Evans}, {Allen}, {Blake}, {Boogert}, {Bourke},
  {Harvey}, {Kessler}, {Koerner}, {Lee}, {Mundy}, {Myers}, {Padgett},
  {Pontoppidan}, {Sargent}, {Stapelfeldt}, {van Dishoeck}, {Young}, \&
  {Young}}]{Evans03}
{Evans}, II, N.~J., {Allen}, L.~E., {Blake}, G.~A., {et~al.} 2003, \pasp, 115,
  965

\bibitem[{{Furlan} {et~al.}(2008){Furlan}, {McClure}, {Calvet}, {Hartmann},
  {D'Alessio}, {Forrest}, {Watson}, {Uchida}, {Sargent}, {Green}, \&
  {Herter}}]{Furlan08}
{Furlan}, E., {McClure}, M., {Calvet}, N., {et~al.} 2008, \apjs, 176, 184

\bibitem[{{Garrod} \& {Herbst}(2006)}]{Garrod06}
{Garrod}, R.~T. \& {Herbst}, E. 2006, \aap, 457, 927

\bibitem[{{Garrod} {et~al.}(2008){Garrod}, {Weaver}, \& {Herbst}}]{Garrod08}
{Garrod}, R.~T., {Weaver}, S.~L.~W., \& {Herbst}, E. 2008, \apj, 682, 283

\bibitem[{{Goldsmith} \& {Langer}(1999)}]{Goldsmith99}
{Goldsmith}, P.~F. \& {Langer}, W.~D. 1999, \apj, 517, 209

\bibitem[{{Hatchell} {et~al.}(2007){Hatchell}, {Fuller}, {Richer}, {Harries},
  \& {Ladd}}]{Hatchell07}
{Hatchell}, J., {Fuller}, G.~A., {Richer}, J.~S., {Harries}, T.~J., \& {Ladd},
  E.~F. 2007, \aap, 468, 1009

\bibitem[{{Helmich} \& {van Dishoeck}(1997)}]{Helmich97}
{Helmich}, F.~P. \& {van Dishoeck}, E.~F. 1997, \aaps, 124, 205

\bibitem[{{Herbst} \& {van Dishoeck}(2009)}]{Herbst09}
{Herbst}, E. \& {van Dishoeck}, E.~F. 2009, \araa, 47, 427

\bibitem[{{J{\o}rgensen} {et~al.}(2002){J{\o}rgensen}, {Sch{\"o}ier}, \& {van
  Dishoeck}}]{Jorgensen02}
{J{\o}rgensen}, J.~K., {Sch{\"o}ier}, F.~L., \& {van Dishoeck}, E.~F. 2002,
  \aap, 389, 908

\bibitem[{{M{\"u}ller} {et~al.}(2001){M{\"u}ller}, {Thorwirth}, {Roth}, \&
  {Winnewisser}}]{Muller01}
{M{\"u}ller}, H.~S.~P., {Thorwirth}, S., {Roth}, D.~A., \& {Winnewisser}, G.
  2001, \aap, 370, L49

\bibitem[{{{\"O}berg} {et~al.}(2013){{\"O}berg}, {Boamah}, {Fayolle}, {Garrod},
  {Cyganowski}, \& {van der Tak}}]{Oberg13}
{{\"O}berg}, K.~I., {Boamah}, M.~D., {Fayolle}, E.~C., {et~al.} 2013, \apj,
  771, 95

\bibitem[{{{\"O}berg} {et~al.}(2008){{\"O}berg}, {Boogert}, {Pontoppidan},
  {Blake}, {Evans}, {Lahuis}, \& {van Dishoeck}}]{Oberg08}
{{\"O}berg}, K.~I., {Boogert}, A.~C.~A., {Pontoppidan}, K.~M., {et~al.} 2008,
  \apj, 678, 1032

\bibitem[{{{\"O}berg} {et~al.}(2011{\natexlab{a}}){{\"O}berg}, {Boogert},
  {Pontoppidan}, {van den Broek}, {van Dishoeck}, {Bottinelli}, {Blake}, \&
  {Evans}}]{Oberg11c}
{{\"O}berg}, K.~I., {Boogert}, A.~C.~A., {Pontoppidan}, K.~M., {et~al.}
  2011{\natexlab{a}}, \apj, 740, 109

\bibitem[{{{\"O}berg} {et~al.}(2010{\natexlab{a}}){{\"O}berg}, {Bottinelli},
  {J{\o}rgensen}, \& {van Dishoeck}}]{Oberg10a}
{{\"O}berg}, K.~I., {Bottinelli}, S., {J{\o}rgensen}, J.~K., \& {van Dishoeck},
  E.~F. 2010{\natexlab{a}}, \apj, 716, 825

\bibitem[{{{\"O}berg} {et~al.}(2009){{\"O}berg}, {Bottinelli}, \& {van
  Dishoeck}}]{Oberg09a}
{{\"O}berg}, K.~I., {Bottinelli}, S., \& {van Dishoeck}, E.~F. 2009, \aap, 494,
  L13

\bibitem[{{{\"O}berg} {et~al.}(2010{\natexlab{b}}){{\"O}berg}, {Qi}, {Fogel},
  {Bergin}, {Andrews}, {Espaillat}, {van Kempen}, {Wilner}, \&
  {Pascucci}}]{Oberg10c}
{{\"O}berg}, K.~I., {Qi}, C., {Fogel}, J.~K.~J., {et~al.} 2010{\natexlab{b}},
  \apj, 720, 480

\bibitem[{{{\"O}berg} {et~al.}(2011{\natexlab{b}}){{\"O}berg}, {van der Marel},
  {Kristensen}, \& {van Dishoeck}}]{Oberg11b}
{{\"O}berg}, K.~I., {van der Marel}, N., {Kristensen}, L.~E., \& {van
  Dishoeck}, E.~F. 2011{\natexlab{b}}, \apj, 740, 14

\bibitem[{{Pickett} {et~al.}(1998){Pickett}, {Poynter}, {Cohen}, {Delitsky},
  {Pearson}, \& {M{\"u}ller}}]{Pickett98}
{Pickett}, H.~M., {Poynter}, R.~L., {Cohen}, E.~A., {et~al.} 1998, \jqsrt, 60,
  883

\bibitem[{{Pontoppidan} {et~al.}(2004){Pontoppidan}, {van Dishoeck}, \&
  {Dartois}}]{Pontoppidan04}
{Pontoppidan}, K.~M., {van Dishoeck}, E.~F., \& {Dartois}, E. 2004, \aap, 426,
  925

\bibitem[{{Sakai} {et~al.}(2010){Sakai}, {Sakai}, {Hirota}, \&
  {Yamamoto}}]{Sakai10}
{Sakai}, N., {Sakai}, T., {Hirota}, T., \& {Yamamoto}, S. 2010, \apj, 722, 1633

\bibitem[{{van der Tak} {et~al.}(2007){van der Tak}, {Black}, {Sch{\"o}ier},
  {Jansen}, \& {van Dishoeck}}]{vanderTak07}
{van der Tak}, F.~F.~S., {Black}, J.~H., {Sch{\"o}ier}, F.~L., {Jansen}, D.~J.,
  \& {van Dishoeck}, E.~F. 2007, \aap, 468, 627

\bibitem[{{Wilking} {et~al.}(2001){Wilking}, {Bontemps}, {Schuler}, {Greene},
  \& {Andr{\'e}}}]{Wilking01}
{Wilking}, B.~A., {Bontemps}, S., {Schuler}, R.~E., {Greene}, T.~P., \&
  {Andr{\'e}}, P. 2001, \apj, 551, 357

\end{thebibliography}

%%%%%%FIGURES%%%%%%%%%%%%%%%%%%%

%%%%%%%TABLES%%%%%%%%%%%%%%%%%%%%

\end{document}